# Significant enhancement of magneto-optical effect in one-dimensional photonic crystals with magnetized epsilon-near-zero defect

Zhiwei Guo[1,†], Feng Wu[1,†], Chunhua Xue[2,*], Haitao Jiang[1,*], Yong Sun[1], Yunhui Li[1], and Hong Chen[1]

*1. Key Laboratory of Advanced Micro-structure Materials, MOE, School of Physics Science and Engineering, Tongji University, Shanghai 200092, China.*

*2. School of Computer Science & Communication Engineering, Guangxi University of Science and Technology, Liuzhou, Guangxi, 545006, China.*

## Abstract

Nonreciprocal (NOR) transmission with magneto-optical materials plays a critical role in a broad range of applications, such as optical isolation, all-optical signal processing, and integrated photonic circuits. The underlying mechanism is that a static magnetic field can break the time-reversal symmetry in the presence of magneto-optical materials. However, the typical NOR devices usually need a large size because the weak magneto-optical activity of materials. Here, our theoretical investigations show that the NOR transmission can be obtained in the one-dimensional photonic crystal with a magnetized epsilon-near-zero (ENZ) defect due to strong field localization in the ENZ medium. The inherent weak magneto-optical activity is significantly enhanced in ENZ medium. In our configuration, the wavelength shift of transmission peaks along two opposite incident directions can be up to 100 times higher than that in the case that the defect layer is a normal bismuth iron garnet. Such results will provide a new degree to design novel magneto-optical devices with small size, and may open novel routes to exploit advanced materials for steering the electromagnetic waves in nano-scale structures.

† These authors contributed equally to this work

* Corresponding author: Email: davies2000@163.com
                        jiang-haitao@tongji.edu

Optical nonreciprocity (ONR), characteristic of light to differentiate between opposite propagating directions, has become an active topic of scientific research [1, 2]. By using magneto-optical (MO) materials such as bismuth iron garnet (BIG) [3] and yttrium iron garnet (YIG) [4], people can realize the corresponding optical nonreciprocal devices including optical isolators [5-9] and circulators [10], which are the key components in optical communication system. However, the ONR effect of nature MO materials is too weak to be widely used in the actual applications. In order to boost the MO effect, diversified optical micro-structures with strong field localization have been put forward. As a kind of typical structure, one-dimensional (1-D) photonic crystals (PCs) with MO materials have been investigated widely [11-14]. However, the MO effect still not strong enough for the significant NOR transmission behaviour, which we will do a detailed analysis in the text.

On the other hand, metamaterials (MMs) composed of sub-wavelength elements have attracted considerable attention as their permittivity and permeability can be designed flexibly [15-17]. As an intriguing class of MMs, epsilon-near-zero (ENZ) MMs have various fantastic applications such as directive emission devices [18], energy squeezing [19], electric field enhancement [20-22], large optical nonlinearity [23-25] and strong optical activity [26] due to their critical permittivity. Recently, Davoyan et al. theoretically proposed the magnetized ENZ MMs and studied the propagation of electromagnetic waves in the bulk and at the surface of a magnetized ENZ medium [27]. Their remarkable finding of the magnetized ENZ medium enables the exploration of new regime about ONR.

In this paper, we show the first study on the ONR transmission in 1-D PCs with magnetized ENZ defect. Our theoretical results show that, in contrast to the conventional MO defect, the magnetized ENZ defect in 1-D PCs will significantly enhance the ONR transmission due to the large field localization in the ENZ medium. Furthermore, we study the dependence of ONR transmission on the incident angle, which helps us optimize the parameters for designing the ONR devices. Moreover, to further evaluate the operation performance quantitatively, we study the influence of the permittivity and the directional external magnetic field on the ONR transmission. Our results provide a pathway to design the novel ONR devices with excellent performance by using ENZ medium.

We start with the study of the wave launching to a 1-D PC with magnetized ENZ defect, as shown schematically in Fig. 1. The structure is denoted by $(AB)^{10}C(AB)^{10}$, where 1-D PC consists of alternative non-magnetized layers of $A(\varepsilon_A, \mu_A)$ and $B(\varepsilon_B, \mu_B)$ with thicknesses of $d_A = 80nm$ and $d_B = 120nm$, respectively. The thickness of unit cell is $d = d_A + d_B$. In addition, a magnetized C layer (marked by pink for seeing in Fig. 1) with thicknesses of $d_C = 50nm$ is inserted into the 1-D PC as a defect layer. Here we select A to be air $(\varepsilon_A = 1, \mu_A = 1)$ and B to be titanium dioxide $(\varepsilon_B = 4.49, \mu_B = 1)$ [28]. Thus far, the magnetized epsilon-near-zero medium can be realized by the magnetized Drude parameters [2] or an unpaired Dirac point [29]. Supposing that the magnetization aligned with an external applied magnetic field is along the y direction, the permittivity $\bar{\varepsilon}_C$ of C is characterized by a dielectric tensor of the following form [14]:

$$\bar{\varepsilon}_c = \begin{pmatrix} \varepsilon_c & 0 & -i\alpha \\ 0 & \varepsilon_c & 0 \\ i\alpha & 0 & \varepsilon_c \end{pmatrix}, \tag{1}$$

where $\varepsilon_c$ is the diagonal component of permittivity tensor and $\alpha$ is the off-diagonal element responsible for the "strength" of MO activity of the medium. The permeability of defect layer is $\mu_C = 1$, and its thickness of defect layer $d_C$.

For such magnetized medium, TM and TE modes (H or E fields polarized along the y direction, respectively) are completely decoupled. Therefore, EM waves will maintain their initial polarization during propagation and $2\times 2$ transfer matrix method can be used. Considering the TM wave propagating in the plane of $xoz$ in our configuration, the electric and magnetic fields in different layer of PC can be written as [11, 12, 14]

$$\begin{cases} E_x(x,z) = (M+N)H_y^+ + (M-N)H_y^- \\ H_y(x,z) = H_0^+ e^{i(k_x x + k_z z - \omega t)} + H_0^- e^{i(k_x x - k_z z - \omega t)} \end{cases}, \quad (2)$$

where $M = -\dfrac{ik_x \alpha}{\varepsilon_0(\varepsilon_c^2 - \alpha^2)\omega}$, $N = \dfrac{k_z \varepsilon_c}{\varepsilon_0(\varepsilon_c^2 - \alpha^2)}$. $k_x$ and $k_z$ are the $x$ and $z$ components of the wave-vector in the plane, respectively. $\omega$ denotes the angular frequency. Superscripts + and – describe the transmitted and reflected waves, respectively.

According to the continuity of the tangential electromagnetic field at the interface, we can obtain the transition matrix between $i^{th}$ layer and $j^{th}$ layer [11, 12, 14]

$$M_{j,i} = T_j^- T_i = \begin{pmatrix} (M_j - N_j)/2N_j & 1/2N_j \\ (M_j + N_j)/2N_j & -1/2N_j \end{pmatrix} \begin{pmatrix} 1 & 1 \\ M_i + N_i & M_i - N_i \end{pmatrix}, \quad (3)$$

For the multilayer structures, the relationship between the input and output magnetic fields can be expressed as

$$\begin{pmatrix} H_{In}^+ \\ H_{In}^- \end{pmatrix} = M_{In} (\prod_{i=1}^{n} P_i M_{i,i+1} P_{i+1}) M_{Out} \begin{pmatrix} H_{Out}^+ \\ H_{Out}^- \end{pmatrix}, \quad (4)$$

where $P_i$ is the usual propagating matrices

$$P_i = \begin{pmatrix} e^{-ik_{zi} d_i} & 0 \\ 0 & e^{ik_{zi} d_i} \end{pmatrix}. \quad (5)$$

Finally, the transmission can be obtained by the following equation

$$T = \left| \dfrac{H_{Out}^+}{H_{in}^+} \right|. \quad (6)$$

Without loss of generality, we consider the nondispersive case. We assume that the parameters of magnetized ENZ material are $\varepsilon_C = 0.1$, $\mu_C = 1$ and $\alpha = 0.06$. Here, for the convenience of comparison, we set the value of $\alpha$ of the magneto-optical coefficients as large as the value of BIG [14]. The most important physical parameter for an ENZ material is field intensity enhancement (FIE), which is defined as [21, 22]

$$FIE = |E_{z2}|^2 / |E_1|^2, \tag{7}$$

where $E_1$ and $E_{z2}$ are the amplitude of incident electric field and the longitudianal electric field in ENZ medium, respectively. The field enhancement for wide layer thickness and angular band is shown in Fig. 2. We can see that the thinner layer thickness and smaller incident angle will lead to the more significant FIE. The values of FIE can be greater than 300. Therefore, combining the strong field location of ENZ layer with the localized defect mode, the MO effect will further enhanced.

Nextly we will study the ONR transmission in the 1-D PC with magnetized ENZ defect. We can find from Fig. 3(a) that the wavelength of defect modes are different for the forward and backward incidences when the light is obliquely incidented to the structure with the incident angle of $\theta = 30^o$. For back illumination, the transmitted peak appears at 707nm (blue dotted line). However, for front illumination, the transmitted peak appears at 690nm (red solid line). The wavelength shift of defect modes between back and front illumination reaches about 17nm which is 8% of the gap width. Moreover, the calculated results are in agreement with the simulations by using COMSOL MULTI-PHYSICS based on the finite-element method, as is shown in Fig. 3(c). In results, the wavelength of transmitted peak of back illumination drops in the bandgap of the front illumination entirely. The results can be validated by simulating the propagation behaviors of electromagnetic waves at wavelength 707nm. For the front illumination, we can see from Fig. 3(e) that the input light (marked by cyan arrow) will be totally reflected (marked by the green arrow).

However, there is a significant change in the event of back illumination. In this situation, the input light will tunnel through the structure (marked by the yellow arrow) with a high transmission, as is illustrated in Fig. 3(g). This transmission is come from the time-reversal symmetry breaking and an ONR transmission behavior occurs as the magnetized ENZ defect is introduced into the 1-D PC.

For comparison, we investigate another case that the magnetized ENZ defect is replaced by the BIG defect. In this situation, all the parameters are same as the previous case except that the permittivity of the defect is changed. The electromagnetic parameters of BIG are $\varepsilon_C = 6.25$ and $\mu_C = 1$ [14]. The transmission spectra of the front and back illuminations are calculated in Fig. 3(b). We can see that the transmitted peaks of the front (the red solid line) and the back illuminations (blue dotted line) nearly overlap at wavelength 633.5 nm. In fact, the wavelength shift of defect modes between back and front illumination just is 0.16nm which is less than 0.1% of the gap width. The simulation in Fig. 3(d) gives the results consistent with those in Fig. 3(b). The results also can be validated by simulating the propagation behaviors of electromagnetic waves. At 633.5 nm for the BIG defect mode, all the input lights tunnel through the structure with high transmission no matter what the direction of the incident light is, as is illustrated in Fig. 3(f) and (h). It means the weak MO activity of BIG defect mode, which usually cannot work for ONR transmission. By comparing the cases of magnetized ENZ and BIG defect, we can find that the magnetized ENZ defect will significantly enhance the ONR transmission about 106 times. This enhancement mechanism is mainly resulted from the large light-matter interaction in the ENZ materials. For conventional MO materials, $|\alpha/\varepsilon_C| \ll 1$, which restrict the MO activity. By contrast, $|\alpha/\varepsilon_C| \sim 1$ for MO ENZ materials, which can enhance the MO activity greatly.

To take into account the influence of incident angle of electromagnetic wave, we calculate the dependence of ONR transmission on the incident angle. For the 1-D PC with magnetized ENZ defect, the wavelengths of transmission peaks for two incident directions will blue shift with an increase of incident angle, which is illustrated in Fig. 4(a). On the other hand, the wavelength shift of defect modes between back and front illumination begins to increase and then decrease with the increase of incident angle $\theta$. The maximum 18nm is obtained at $\theta = 33^o$, which is shown in Fig. 4(c). Although the PC with magnetized BIG defect also have the similar property (see Fig. 4(d)), the ONR transmission is too weak to observation (see Fig. 4(b)). The maximum of the wavelength shift just is 0.16nm, which is far less than the case of magnetized ENZ defect. The comparison between magnetized ENZ and BIG defect layers further demonstrate the significant enhancement of ONR transmission of magnetized ENZ media from a wide range of incident angle. Lastly, we focus on the influence of varied permittivity of the defect layer on the transmission for both incident directions. Here we define $\delta T = T_F - T_B$ as the difference of transmission between both directions, where $T_F$ and $T_B$ represent the transmittance of front and back illumination, respectively. In this case other parameters are fixed as $\theta = 33^o$, $d_C = 50nm$ and $\alpha = 0.06$. The difference of transmission between both directions is illustrated in Fig. 5, in which the red, blue and black areas indicate $\delta T = 1$, $\delta T = -1$ and $\delta T = 0$. Undoubtedly, the absolute value of the transmission difference represents the performance of ONR transmission. When the value of $\varepsilon_C$ is near to zero, as is shown in Fig. 5, the performance of ONR is apparent as the wavelength difference between two transmission peaks is large. However, with the increase of the permittivity of the defect layer, the ONR effect gradually decreases. The ONR effect disappears entirely when $\varepsilon_C > 0.4$. The results show that the performance of ONR transmission is sensitive to the permittivity of defect layer. A properly small value of $\varepsilon_C$ contributes to the realization of ONR transmission with excellent performance.

# CONCLUTION

In conclusion, beyond the only resonant effect and the increase of the "strength" of MO activity, our investigations show that magnetized ENZ defect layer will significantly enhance the magneto-optical effect (NOR transmission) in 1-D PCs. Our results not only may open novel routes to exploit advanced materials for steering the electromagnetic waves in nano-scale structures, but also may have various potential applications in optical isolators, circulators, sensors and switches.

# ACKNOWLEDGMENT

This work is supported by the National Key R&D Program of China (No. 2016YFA0301101); by the National Nature Science Foundation of China (NSFC) (Grant Nos. 61661007, 11774261, 11474220, and 61621001); Guangxi Natural Science Foundation (No. 2016GXNSFAA380198); Science Foundation of Shanghai (No. 17ZR1443800).

**Figure Captions**

FIG. 1. (Color online) Schematic of ONR transmission in the 1-D PC with magnetized defect. The incident light from the left (shown by red) can pass through the PC, while the incident light from the right (shown by blue) is totally reflected.

FIG. 2. (Color online) The FIE of ENZ media as a function of incident angle and thickness of layer, assuming a TM-polarized plane wave incident at 400 THz.

FIG. 3. (Color online) (a, b) Calculated transmission spectra and (c, d) simulated transmission behaviors of front (the red trace) and back illumination (the blue trace) for the 1-D PC with (a, c) magnetized ENZ defect and (b, d) BIG defect. (e, g) Magnetic-field pattern of the 1-D PC with magnetized ENZ defect at $\lambda = 707 nm$. (f, h) Magnetic-field pattern of the 1-D PC with BIG defect at $\lambda = 633.5 nm$. Interfaces between the external air and the 1-D PC and positions of the defect layer are marked by green and yellow lines, respectively.

FIG. 4. (Color online) (a, b) Dependence of wavelength of transmission peak on the incident angle in the PCs with magnetized ENZ and BIG defect layer for two opposite directions. (c, d) Wavelength difference of transmission peaks along two opposite incident directions in two PC structures with different defect layer.

FIG. 5. (Color online) $\delta T$ for the variant permittivity of defect layer. Other parameters are $\theta = 33^o$, $d_C = 50 nm$ and $\alpha = 0.06$.

**Figures**

FIG.1

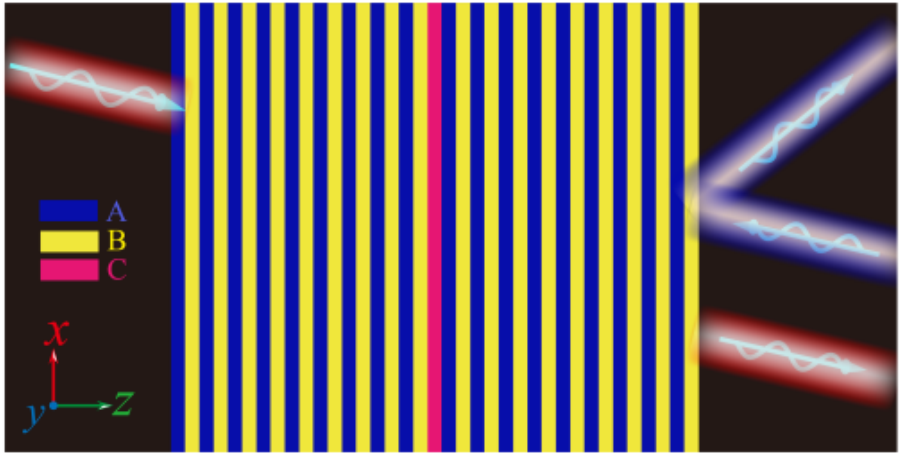

FIG.2

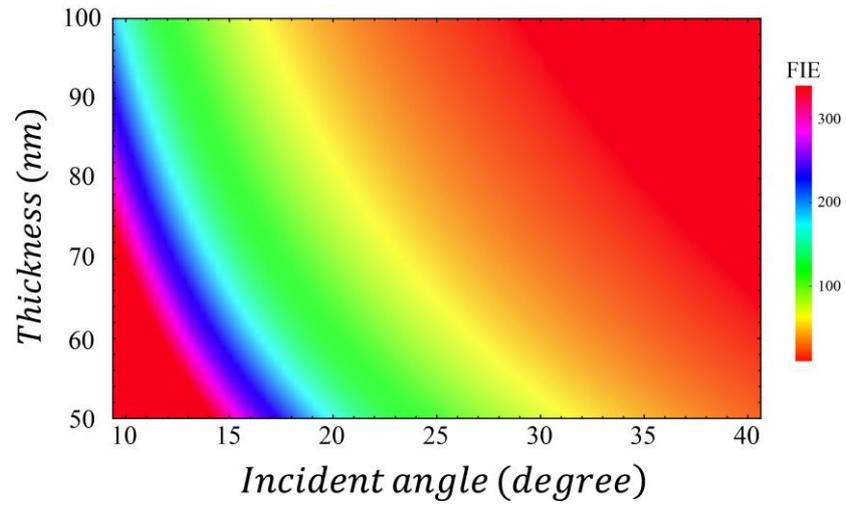

FIG.3

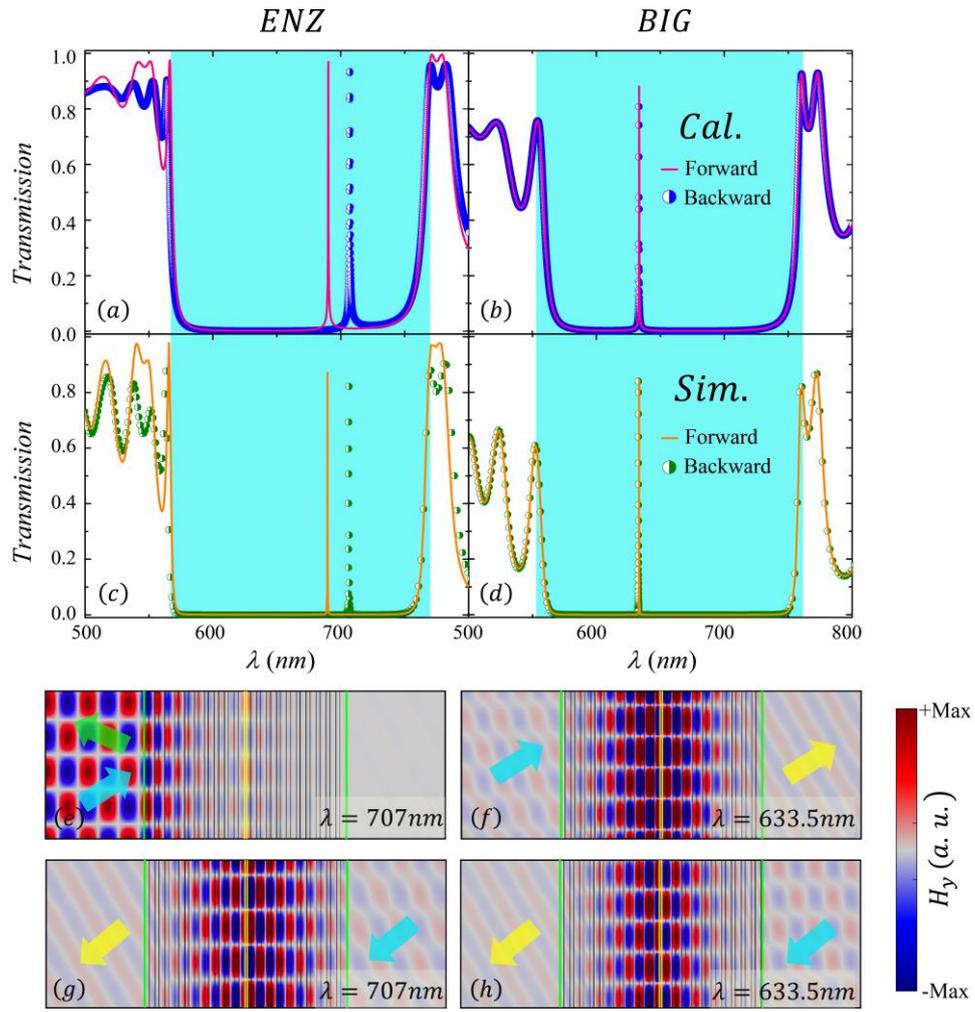

FIG.4

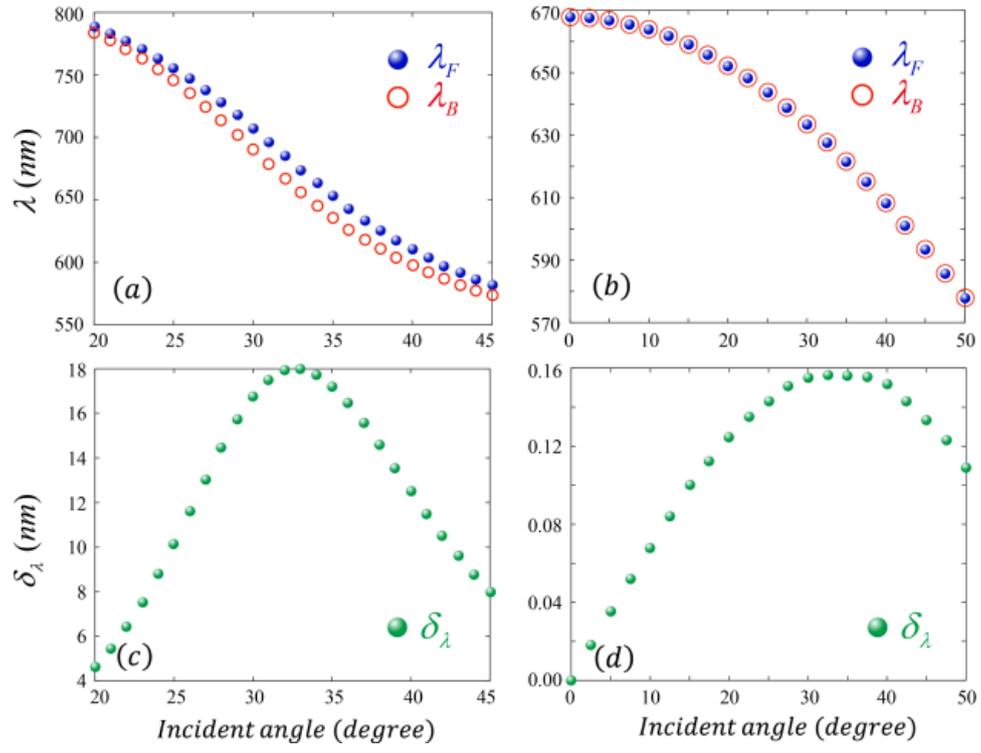

FIG.5

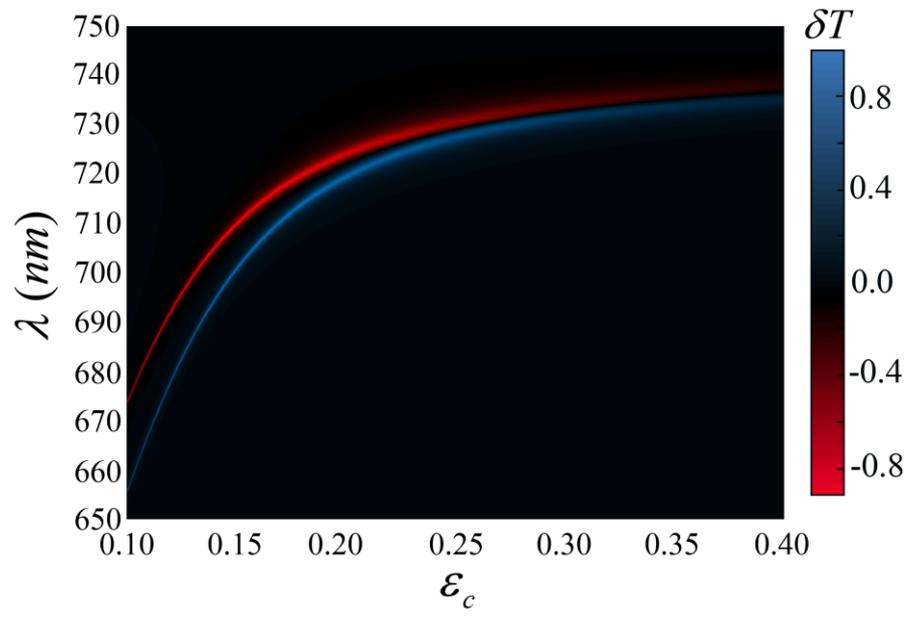